\documentclass[twocolumn,showpacs,showkeys,nofootinbib,prb]{revtex4}
\usepackage{graphicx}
\usepackage{soul}
\usepackage{color}
\usepackage{amsmath}
\usepackage{stackrel}
\begin{document}
\title{
SQUID Metamaterials on a Lieb lattice: From flat-band to nonlinear localization}  
\author{N. Lazarides$^{1,2,3,4}$, G. P. Tsironis$^{1,2,3,4}$}
\affiliation{
$^{1}$Department of Physics, University of Crete, P. O. Box 2208, 71003 Heraklion, 
      Greece; \\
$^{2}$Institute of Electronic Structure and Laser, Foundation for Research and 
      Technology--Hellas, P.O. Box 1527, 71110 Heraklion, Greece \\
$^{3}$National University of Science and Technology "MISiS", Leninsky prosp. 4, Moscow, 
      119049, Russia; \\ 
$^{4}$Department of Physics, School of Science and Technology, Nazarbayev University,
      53 Kabanbay Batyr Ave., Astana 010000, Kazakhstan
}
\date{\today}
\begin{abstract}
The dynamic equations for the fluxes through the SQUIDs that form a two-dimensional 
metamamaterial on a Lieb lattice are derived, and then linearized around zero flux to 
obtain the {\em linear frequency spectrum} according to the standard procedure. That 
spectrum, due to the Lieb lattice geometry, possesses a frequency band structure 
exhibiting two characteristic features; two dispersive bands, which form a Dirac
cone at the corners of the first Brillouin zone, and a flat band crossing the Dirac 
points. It is demonstrated numerically that localized states can be excited in the 
system when it is initialized with single-site excitations; depending on the amplitude
of those initial states, the localization is either due to the flat-band or to nonlinear
effects. Flat-band localized states are formed in the nearly linear regime, while 
localized excitations of the discrete breather type are formed in the nonlinear regime.
These two regimes are separated by an intermediate turbulent regime for which no 
localization is observed. Notably, initial single-site excitations of only edge SQUIDs 
of a unit cell may end-up in flat-band localized states; no such states are formed for 
initial single-site excitations of a corner SQUID of a unit cell. The degree of 
localization of the resulting states is in any case quantified using well-established 
measures such as the energetic participation ratio and the second moment.
\end{abstract}
\pacs{63.20.Pw, 11.30.Er, 41.20.-q, 78.67.Pt}
\keywords{SQUID metamaterials, Lieb lattice, flat dispersion bands, localized flat-band 
states, Hamiltonian systems, Discrete breathers} 
\maketitle
\section{Introduction}
Considerable research effort has focused the last two decades in the investigation and 
developement of artificial mediums or {\em metamaterials}, which exhibit properties 
not found in natural materials \cite{Smith2004,Yen2004,Linden2004,Linden2006,Shalaev2007,
Litchinitser2008}. After the development of active, tunable, and nonlinear metamaterials
\cite{Boardman2010,Lapine2014}, those artificial mediums are expected to have a strong 
impact across the entire range of technologies where electromagnetic radiation is used. 
Moreover, they may provide a flexible platform for modeling and mimicking fundamental 
physical effects \cite{Zheludev2010,Zheludev2011,Zheludev2012}. An imporant class of 
metamaterials is that of superconducting ones \cite{Anlage2011,Jung2014}, and in 
particular those comprising Superconducting QUantum Interference Devices (SQUIDs). 
The idea of a metamaterial consisting of SQUIDs was theoretically introduced about a 
decade ago both in the quantum \cite{Du2006} and the classical regimes 
\cite{Lazarides2007}. 

\begin{figure}[!h]
\includegraphics[angle=0, width=0.9 \linewidth]{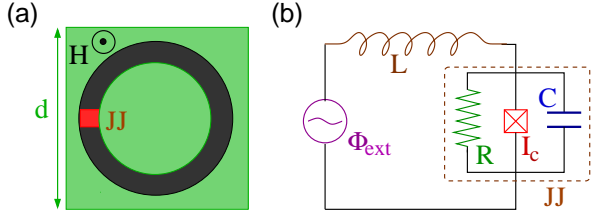}
\caption{(Color online)
Schematic of a SQUID (a) and its equivalent electrical circuit (b).
\label{fig1}
}
\end{figure}

The simplest version of a SQUID consists of a superconducting ring interrupted by a 
Josephson junction \cite{Josephson1962}, as shown schematically in Fig. \ref{fig1}. 
The SQUIDs are highly nonlinear devices, exhibiting strong resonant response to applied 
magnetic fields. SQUID metamaterials in one and two dimensions have been realized and 
investigated in the laboratory, and they were found to exhibit novel properties such as 
negative diamagnetic permeability \cite{Jung2013,Butz2013a}, broad-band tunability 
\cite{Butz2013a,Trepanier2013}, self-induced broad-band transparency \cite{Zhang2015}, 
as well as dynamic multistability and switching \cite{Jung2014b}, among others. 
Some of these properties, i.e., dynamic multistability and tunability, have been also
revealed in numerical simulations \cite{Lazarides2013b,Tsironis2014b}. 
Moreover, nonlinear localization \cite{Lazarides2008a} and the emergence of 
counter-intuitive dynamic states referred to as {\em chimera states} in current 
literature \cite{Lazarides2015b,Hizanidis2016a} have been demonstrated numerically 
in SQUID metamaterial models.

The notion of metamaterials implies the freedom to engineer not only the properties of
the individual "particles" or devices which play the role of "atoms" in an artificial 
medium, but also their arrangement in space, i.e., the type of the lattice. 
Remarkably, some specific lattice geometries such as those of {\em Lieb} or 
{\em Kagom{\'e}} lattices give rise to novel and potentially useful band structures. 
The former is a square-depleted (line-centered tetragonal) lattice, described by three 
sites in a square unit cell as illustrated in Fig. \ref{fig2}. It is characterized by 
a band structure featuring Dirac cones intersected by a topological flat band. 
Localization on flat-bands has been extensively investigated in relatively simple 
lattice models \cite{Danieli2015,Khomeriki2016}, even in the presence of disorder 
\cite{Leykam2013,Leykam2017}. Superpositions of flat-band modes and their stability 
have been also investigated in rhombic nonlinear optical waveguide arrays 
\cite{Maimistov2017}. The Lieb lattice was first introduced in the context of 
photonics in Ref. \cite{Leykam2012}. Recently, photonic Lieb lattices have been 
experimentally realized and the existence of localized flat-band modes has been 
reported \cite{Mukherjee2015a,Vicencio2015}. The world of electronic flat-band systems 
has been reviewed in a recent article \cite{LiuZheng2014}. Moreover, electronic Lieb 
lattices have been experimentally realized and characterized \cite{Slot2017}. 
Here, a SQUID metamaterial on a Lieb lattice is considered, in which each site is 
occupied by a SQUID. In each unit cell, two of the SQUIDs (indicated in red and blue) 
are neighbored by two other SQUIDs. The third SQUID in the unit cell (black) has four  
neighbors. In what follows, these SQUIDs will be referred to as edge SQUIDs (red and 
blue) and corner (black) SQUID, respectively. 

In the following, the dynamical equations for the fluxes through the SQUIDs and the 
linear frequency spectrum are obtained for a SQUID Lieb metamaterial (SLiMM). Using 
numerical simulations, the generation of {\em localized flat band states} when a 
single edge SQUID is initially excited at low amplitude, is demonstrated. No 
flat-band localization is observed when single corner SQUIDs are initially excited at 
low amplitudes, in agreement with the experiments in optical Lieb lattices 
\cite{Guzman-Silva2014,Mukherjee2015a,Vicencio2015}. For high-amplitude initial 
excitations of either a corner or an edge SQUID, nonlinear localization of the 
discrete breather type is observed \cite{Flach2008a}. The cross-over between flat-band 
and nonlinear localization is explored; the two regimes are clearly separated by an 
intermediate, no-localization regime. Thus, flat-band localized states cannot be 
continuated into the nonlinearly localized ones of the discrete breather or discrete
soliton type, as it has been demonstrated for discrete nonlinear Schr{\"o}dinger type 
models of various flat-band lattices and ribbons
\cite{Vicencio2013,Leykam2013,Johansson2015,Belicev2015,Lopez-Gonzalez2016,Gligoric2016}.

\section{Flux Dynamics}
\begin{figure}[!h]
\includegraphics[angle=0, width=0.9 \linewidth]{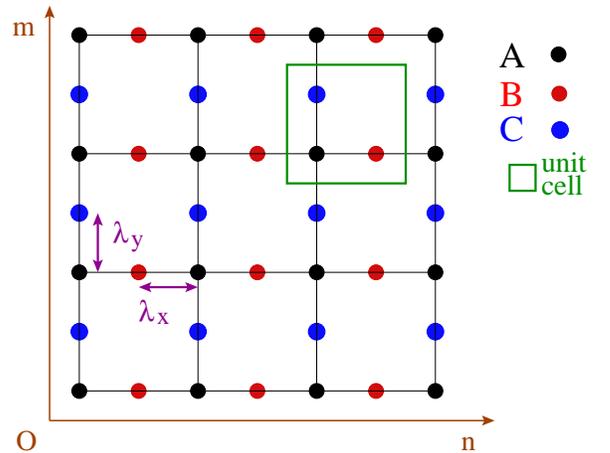}
\caption{(Color online)
Schematic of a Lieb lattice in which each site is occupied by a SQUID. The three 
sublattices are indicated in black (corner SQUIDs), red (edge SQUIDs), and blue (edge 
SQUIDs) color. The nearest-neighbor couplings along the horizontal ($\lambda_x$) and 
vertical ($\lambda_y$) directions and the unit cell are also indicated.   
\label{fig2}
}
\end{figure}

Consider the Lieb lattice of Fig. \ref{fig2}, in which each site is occupied by a 
SQUID. That SLiMM can be regarded as the combination of three sublattices colored as 
blue, red, and black. All the SQUIDs are identical, and they are magnetically coupled 
to their nearest neighbors through their mutual inductances. In order to derive the 
dynamic equations for the fluxes through the SQUIDs of the SLiMM, we first write the 
flux-balance relations for all SQUIDs
\begin{eqnarray}
  \Phi_{n,m}^A = \Phi_{ext} +L \left\{
  I_{n,m}^A +\lambda_x \left[ I_{n-1,m}^B +I_{n,m}^B \right] \right. 
\nonumber \\ 
     \left. +\lambda_y \left[ I_{n,m-1}^C +I_{n,m}^C \right] \right\} , 
\nonumber \\
\label{1}
  \Phi_{n,m}^B = \Phi_{ext} +L \left\{
  I_{n,m}^B +\lambda_x \left[ I_{n,m}^A +I_{n+1,m}^A \right] \right\} , \\
  \Phi_{n,m}^C = \Phi_{ext} +L \left\{
  I_{n,m}^C +\lambda_y \left[ I_{n,m}^A +I_{n,m+1}^A \right] \right\} , 
\nonumber
\end{eqnarray}
where $I_{n,m}^k$ is the current in the SQUID of the $(n,m)$th unit cell of kind $k$ 
($k=A, B, C$), $\Phi_{ext}$ is the applied (external) flux, 
and $\lambda_x =M_x /L$ ($\lambda_y =M_y /L$) is the coupling coefficient along the 
horizontal (vertical) direction, with $M_x$ ($M_y$) being the corresponding mutual 
inductance between neighboring SQUIDs and $L$ the self-inductance of each SQUID. 
The current in each SQUID is given by the resistively and capacitively shunted 
junction (RCSJ) model \cite{Likharev1986}, as
\begin{equation}
\label{2}
  -I_{n,m}^k =C\frac{d^2 \Phi_{n,m}^k}{dt^2} +\frac{1}{R} \frac{d \Phi_{n,m}^k}{dt}
   +I_c \sin\left( 2\pi \frac{\Phi_{n,m}^k}{\Phi_0} \right) ,
\end{equation}
where $R$ is the quasiparticle resistance through the Josephson junction of each SQUID,
$C$ is the capacitance of each SQUID, and $I_c$ is the critical current of the 
Josephson junction of each SQUID. Then Eqs. (\ref{1}) are inverted to obtain the 
currents $I_{n,m}^k$ as functions of the fluxes $\Phi_{n,m}^k$. By substitution of the 
obtained currents back into Eqs. (\ref{1}), and neglecting all the terms which 
are proportional to $\lambda_x^a \lambda_y^b$ with $a+b >1$, we get
\begin{eqnarray}  
  L I_{n,m}^A = \Phi_{n,m}^A -\lambda_x \left( \Phi_{n,m}^B +\Phi_{n-1,m}^B \right)
\nonumber \\
             -\lambda_y \left( \Phi_{n,m}^C +\Phi_{n,m-1}^C \right) -\Phi_{eff}^A , 
\nonumber \\
\label{3}
  L I_{n,m}^B =\Phi_{n,m}^B -\lambda_x \left( \Phi_{n,m}^A +\Phi_{n+1,m}^A \right)
  -\Phi_{eff}^B , \\
  L I_{n,m}^C =\Phi_{n,m}^C -\lambda_y \left( \Phi_{n,m}^A +\Phi_{n,m+1}^A \right)
  -\Phi_{eff}^C , \nonumber
\end{eqnarray}
where $\Phi_{eff}^A =[1-2(\lambda_x +\lambda_y)] \Phi_{ext}$, 
$\Phi_{eff}^B =( 1-2 \lambda_x ) \Phi_{ext}$, 
and $\Phi_{eff}^C =( 1-2 \lambda_y ) \Phi_{ext}$ are the "effective" external fluxes.
Combining Eqs. (\ref{2}) and (\ref{3}) we get
\begin{eqnarray}
  L C \frac{d^2 \Phi_{n,m}^A}{dt^2} +\frac{L}{R} \frac{d \Phi_{n,m}^A}{dt}
   +L I_c \sin\left( 2\pi \frac{\Phi_{n,m}^A}{\Phi_0} \right) +\Phi_{n,m}^A 
\nonumber \\
  = \lambda_x \left( \Phi_{n,m}^B +\Phi_{n-1,m}^B \right)
   +\lambda_y \left( \Phi_{n,m}^C +\Phi_{n,m-1}^C \right) 
   +\Phi_{eff}^A , 
\nonumber \\
\label{4}
  L C \frac{d^2 \Phi_{n,m}^B}{dt^2} +\frac{L}{R} \frac{d \Phi_{n,m}^B}{dt}
   +L I_c \sin\left( 2\pi \frac{\Phi_{n,m}^B}{\Phi_0} \right) +\Phi_{n,m}^B 
\nonumber \\
  = \lambda_x \left( \Phi_{n,m}^A +\Phi_{n+1,m}^A \right) +\Phi_{eff}^B , 
 \\ 
  L C \frac{d^2 \Phi_{n,m}^C}{dt^2} +\frac{L}{R} \frac{d \Phi_{n,m}^C}{dt}
   +L I_c \sin\left( 2\pi \frac{\Phi_{n,m}^C}{\Phi_0} \right) +\Phi_{n,m}^C 
\nonumber \\
  = \lambda_y \left( \Phi_{n,m}^A +\Phi_{n,m+1}^A \right) +\Phi_{eff}^C . 
\nonumber
\end{eqnarray}

Using the relations 
\begin{equation}
\label{5}
  \tau =\omega_{LC} t , \qquad
  \phi_{n,m}^k=\frac{\Phi_{n,m}^k}{\Phi_0} , \qquad 
  \phi_{ext}=\frac{\Phi_{ext}}{\Phi_0} ,
\end{equation}
where $\omega_{LC} = {1}/{\sqrt{LC}}$ is the inductive-capacitive SQUID frequency,
the dynamic equations for the fluxes through the SQUIDs can be written in the 
normalized form
\begin{eqnarray}
  \ddot{\phi}_{n,m}^A+ \gamma \dot{\phi}_{n,m}^A
   +\beta \sin\left( 2\pi \phi_{n,m}^A \right) +\phi_{n,m}^A 
\nonumber \\
  = \lambda_x \left( \phi_{n,m}^B +\phi_{n-1,m}^B \right) 
   +\lambda_y \left( \phi_{n,m}^C +\phi_{n,m-1}^C \right) 
   +\phi_{eff}^A , 
\nonumber \\
\label{6}
  \ddot{\phi}_{n,m}^B +\gamma \dot{\phi}_{n,m}^B
   +\beta \sin\left( 2\pi \phi_{n,m}^B \right) +\phi_{n,m}^B 
\nonumber \\
  = \lambda_x \left( \phi_{n,m}^A +\phi_{n+1,m}^A \right) +\phi_{eff}^B , 
\\
  \ddot{\phi}_{n,m}^C +\gamma \phi_{n,m}^C
   +\beta \sin\left( 2\pi \phi_{n,m}^C \right) +\phi_{n,m}^C 
\nonumber \\
  = \lambda_y \left( \phi_{n,m}^A +\phi_{n,m+1}^A \right) +\phi_{eff}^C , 
\nonumber
\end{eqnarray}
where 
\begin{equation}
\label{7}
  \beta =\frac{L\, I_c}{\Phi_0} ~~~{\rm and }~~~\gamma= \omega_{LC} \frac{L}{R},   
\end{equation}
is the SQUID parameter and the loss coefficient, respectively, and $\phi_{eff}^k$ are
the normalized effective fluxes. The overdots on $\phi_{n,m}^k$ denote differentiation 
with respect to the normalized temporal variable $\tau$. 
The values of the fluxes through the SQUIDs $\phi_{n,m}^k$ generally depend on $k$. 
Suppose that $\gamma=0$ and $\phi_{ext} =0$, and that Eqs. (\ref{6}) are initialized 
with a low amplitude homogeneous excitation, i.e., with $\phi_{n,m}^k =c$ for any $n$, 
$m$, and $k$ ($c \ll 1$ is a constant). After integrating Eqs. (\ref{6}) in time 
assuming periodic boundary conditions, at the steady state, the fluxes through the 
SQUIDs of the same kind will be the same. However, the fluxes through the SQUIDs of 
different kind will be different. This is due to the Lieb lattice geometry and the 
(generally) different values of $\lambda_x$ and $\lambda_y$, since the flux through 
a particular SQUID of the SLiMM depends not only on the self-induced one, but also on 
the fluxes from the SQUIDs to which that particular SQUID is coupled (four for $A$ 
SQUIDs and two for $B$ and $C$ SQUIDs). Moreover, the coupling between SQUIDs is 
proportional to the coefficients $\lambda_x$ or $\lambda_y$. Note that for isotropic 
coupling, $\lambda_x =\lambda_y$, the fluxes through the SQUIDs of kind $B$ and $C$ 
are the same but different than those through the SQUIDs of kind $A$ 
($\phi_{n,m}^B =\phi_{n,m}^C \neq \phi_{n,m}^A$).  

In the following, we are concerned about energy-conserving SLiMMs, i.e., about the 
Hamiltonian version of SQUID Lieb metamaterials, and thus we set $\gamma=0$ and 
$\phi_{ext}=0$ into Eqs. (\ref{6}). 

\section{Linear Frequency Spectrum}
Without losses and driving forces, Eqs. (\ref{6}) are linearized using the relation
$\beta \, \sin \left( 2\pi \phi_{n,m}^k \right) \simeq \beta_L \, \phi_{n,m}^k$, where 
$\beta_L =2 \pi \beta$. Thus we get  
\begin{eqnarray}
  \ddot{\phi}_{n,m}^A +\Omega_{SQ}^2 \phi_{n,m}^A 
  =\lambda_x \left( \phi_{n,m}^B +\phi_{n-1,m}^B \right) 
\nonumber \\
  +\lambda_y \left( \phi_{n,m}^C +\phi_{n,m-1}^C \right) ,
\nonumber \\
\label{8}
  \ddot{\phi_{n,m}^B} +\Omega_{SQ}^2 \phi_{n,m}^B 
  =\lambda_x \left( \phi_{n,m}^A +\phi_{n+1,m}^A \right) , 
\\
  \ddot{\phi}_{n,m}^C +\Omega_{SQ}^2 \phi_{n,m}^C 
  =\lambda_y \left( \phi_{n,m}^A +\phi_{n,m+1}^A \right) , 
\nonumber
\end{eqnarray} 
where $\Omega_{SQ} =\sqrt{ 1 +\beta_L }$ is the resonance frequency of individual 
SQUIDs in the linear limit. In order to obtain the linear frequency spectrum,
we substitute into the linearized Eqs. (\ref{8}) the plane wave solution 
\begin{equation}
\label{9}
  \phi_{n,m}^k = {\cal F}_k \exp[i (\Omega \tau -\kappa_x n -\kappa_y m)],
\end{equation}
where $\kappa_x$ and $\kappa_y$ are the $x$ and $y$ components of the two-dimensional, 
normalized wavevector $\bf \kappa$, and $\Omega =\omega / \omega_{LC}$ is the 
normalized frequency. After some calculations we get
\begin{eqnarray}
  \left( \Omega_{SQ}^2 -\Omega^2 \right) {\cal F}_A 
 -\lambda_x \left( 1 +e^{+i \kappa_x} \right) {\cal F}_B
\nonumber \\
 -\lambda_y \left( 1 +e^{+i \kappa_y} \right) {\cal F}_C =0, 
\nonumber \\
\label{10}
 -\lambda_x \left( 1 +e^{-i \kappa_x} \right) {\cal F}_A
 +\left( \Omega_{SQ}^2 -\Omega^2 \right) {\cal F}_B =0, 
\\
 -\lambda_y \left( 1 +e^{-i \kappa_y} \right) {\cal F}_A
 +\left( \Omega_{SQ}^2 -\Omega^2 \right) {\cal F}_C =0 . 
\nonumber 
\end{eqnarray} 
In order to obtain nontrivial solutions for the amplitudes ${\cal F}_k$ of the 
stationary problem Eqs. (\ref{10}), its determinant ${\cal D}$ should be equal 
to zero, i.e., 
\begin{eqnarray}
\label{11}
  {\cal D}=\left( \Omega_{SQ}^2 -\Omega^2 \right) 
           \left\{ \left( \Omega_{SQ}^2 -\Omega^2 \right)^2
\right. \nonumber \\ \left.
          -4 \left[ \lambda_x^2 \cos^2 \left(\frac{\kappa_x}{2} \right)
                   +\lambda_y^2 \cos^2 \left(\frac{\kappa_y}{2} \right) 
                \right] \right\} =0 .
\end{eqnarray}
\begin{figure}[!t]
\includegraphics[angle=0, width=0.9 \linewidth]{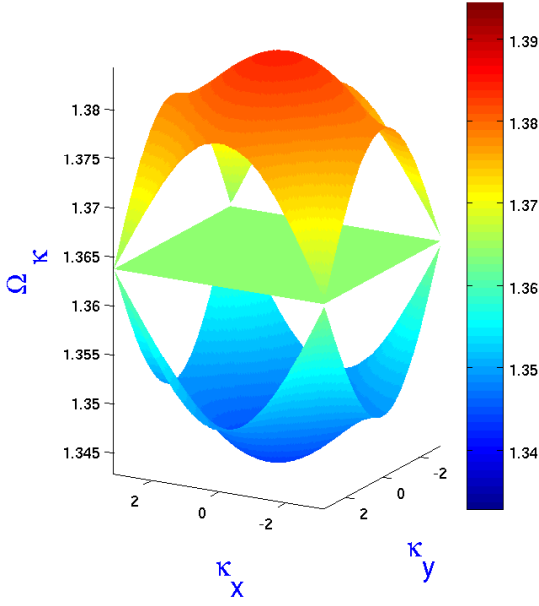} 
\caption{(Color online)
The linear frequency spectrum $\Omega_{\bf \kappa} ({\bf \kappa})$ of the SQUID Lieb 
metamaterial for $\beta=0.86$, and $\lambda_x =\lambda_y =-0.02$. The flat-band 
frequency is $\Omega_{FB} =\Omega_{SQ} \simeq 1.364$. 
\label{fig3}
}
\end{figure}
Solving Eq. (\ref{11}) for $\Omega \equiv \Omega_{\bf \kappa}$, we get
\begin{eqnarray}
\label{12}
  \Omega_{\bf \kappa} &=& \Omega_{SQ}, \\
\label{13}
  \Omega_{\bf \kappa} &=& \sqrt{ \Omega_{SQ}^2 \pm 2 \sqrt{ 
                    \lambda_x^2 \cos^2 \left(\frac{\kappa_x}{2} \right)
                   +\lambda_y^2 \cos^2 \left(\frac{\kappa_y}{2} \right) } } ,  
\end{eqnarray} 
where only positive frequencies are considered. Eqs. (\ref{12}) and (\ref{13}) provide 
the {\em linear frequency spectrum of the SLiMM}. Thus, the Lieb lattice geometry 
possesses a frequency band structure exhibiting two characteristic features as can be 
observed in Fig. \ref{fig3}: two dispersive bands, which form a Dirac cone at the 
corners of the first Brillouin zone (for $\kappa_x =\kappa_y =\pm \pi$), and a flat 
band crossing the Dirac points. It is well-established that Dirac cones give rise to 
peculiar topological properties \cite{Weeks2010} and unusual behavior in general, 
such as effectively massless fermions, etc. Note that the flat-band frequency 
$\Omega_{FB}$ is equal to the resonance frequency of individual SQUIDs $\Omega_{SQ}$, 
i.e., $\Omega_{FB} =\Omega_{SQ}$. The maximum and minimum frequencies of the spectrum 
are obtained from Eq. (\ref{13}) for $\kappa_x =\kappa_y =0$ and 
$\kappa_x =\kappa_y =\pi/2$, respectively, as
\begin{equation}
\label{14}
  \Omega_{\stackrel{max}{min}} 
                    =\sqrt{ \Omega_{SQ}^2 \pm 2 \sqrt{ \lambda_x^2 +\lambda_y^2 } } .
\end{equation} 
Since $|\lambda_x|, |\lambda_y| \ll 1$, the bandwidth of the spectrum is approximatelly
$\Delta \Omega \simeq 2 \sqrt{ \lambda_x^2 +\lambda_y^2 } / \Omega_{SQ}$.
For example, for the parameters of Fig. \ref{fig3} we have $\Omega_{min} \simeq 1.343$, 
$\Omega_{max} \simeq 1.384$, and $\Delta \Omega \simeq 0.04$. 
We also note that the flat band is an intrinsic property of this lattice in the 
nearest-neighbor coupling limit and thus it is not destroyed by any anisotropy (i.e., 
when $\lambda_x \neq \lambda_y$).

\begin{figure}[!t]
\includegraphics[angle=0, width=0.9 \linewidth]{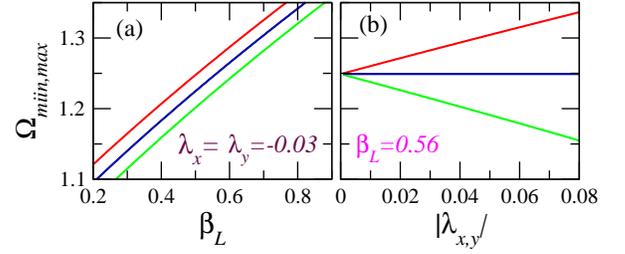} 
\caption{(Color online)
Minimum $\Omega_{min}$ (green), maximum $\Omega_{max}$ (red), and flat-band 
frequency $\Omega_{FB}$ (blue), as a function of (a) the SQUID parameter $\beta_L$
for $\lambda_x =\lambda_y =-0.03$, and 
(b) the coupling coefficients $\lambda_x =\lambda_y$ for $\beta_L =0.56$.
\label{fig4}
}
\end{figure}

The dependence of the extremal frequencies $\Omega_{min,max}$ and the flat-band 
frequency $\Omega_{FB}$ on the parameters $\beta_L$ and $\lambda_x, \lambda_y$ is 
shown in Fig. \ref{fig4}. In Fig. \ref{fig4}a all curves increase linearly with 
increasing $\beta_L$ while the bandwidth remains practically constant. In Fig. 
\ref{fig4}b the bandwidth increases with increasing $\lambda_x =\lambda_y$ while 
$\Omega_{FB}$ remains the same.   

\section{Numerics and Localization Measures}
The dynamic equations for the fluxes through the SQUIDs, Eqs. (\ref{6}), without 
losses and external forcing ($\gamma=0$ and $\phi_{ext} =0$), can be derived as the 
Hamilton's equations from the Hamiltonian function
\begin{equation}
\label{15}
  H =\sum_{n,m} H_{n,m} ,
\end{equation}
where the energy (Hamiltonian) density $H_{n,m}$, defined as the energy per unit
cell, is given by 
\begin{eqnarray}
\label{16}
  H_{n,m} =\sum_{k} \left\{ \frac{\pi}{\beta} \left[ 
           \left( q_{n,m}^k \right)^2 +\left( \phi_{n,m}^k \right)^2 \right]
  -\cos\left( 2\pi \phi_{n,m}^k \right) \right\}
\nonumber \\
  -\frac{\pi}{\beta} \{
   \lambda_x [  \phi_{n,m}^A \phi_{n-1,m}^B  +2 \phi_{n,m}^A \phi_{n,m}^B 
               +\phi_{n,m}^B \phi_{n+1,m}^A ]
\nonumber \\
   +\lambda_y [ \phi_{n,m}^A \phi_{n,m-1}^C 
               +2 \phi_{n,m}^A \phi_{n,m}^C +\phi_{n,m}^C \phi_{n,m+1}^A ] \} ,
~~~
\end{eqnarray}
where $q_{n,m}^k =\frac{ d\phi_{n,m}^k }{d \tau}$ is the normalized instantaneous 
voltage across the Josephson junction of the SQUID in the $(n,m)$th unit cell of kind 
$k$. Both $H$ and $ H_{n,m}$ are normalized to the Josephson energy, $E_J$. The total 
energy $H$, given by Eqs. (\ref{15}) and (\ref{16}), remains constant in time.

Eqs. (\ref{6}) with $\gamma=0$ and $\Phi_{ext} =0$ are integrated in time using the 
second order symplectic 
St{\"o}rmer-Verlet scheme \cite{Hairer2003}, which preserves the total energy $H$ to 
a prescribed accuracy which is a function of the time-step $h$. In the flux - voltage 
variables, that scheme reads \cite{Hairer2003}
\begin{eqnarray}
\vec{\phi}_{n+\frac{1}{2}}^k =\vec{\phi}_{n}^k +\frac{h}{2} \vec{q}_{n}^k , \nonumber \\
\label{17}
\vec{q}_{n+1}^k =\vec{q}_{n}^k -h \, H_{\vec{\phi}^k} (\vec{\phi}_{n+\frac{1}{2}}^k) , \\
\vec{\phi}_{n+1}^k =\vec{\phi}_{n+\frac{1}{2}}^k +\frac{h}{2} \vec{q}_{n+1}^k , \nonumber
\end{eqnarray}
where $\vec{\phi}^k$ and $\vec{q}^k$ are $N-$dimensional vectors ($N=N_x N_y$) containing 
the fluxes and the voltages for the SQUIDs of kind $k$ ($k=A, B, C$), and 
$H_{\vec{\phi}^k} \equiv \nabla_{\vec{\phi}^k} H$ denotes the column vector of partial
derivatives of the Hamiltonian with respect to $\vec{\phi}^k$, i.e., 
\[H_{\vec{\phi}^k} =\left[ \frac{\partial H}{\partial \phi_1^k},~ 
                           \frac{\partial H}{\partial \phi_2^k},~
                           \frac{\partial H}{\partial \phi_3^k},~ ...,~ 
                           \frac{\partial H}{\partial \phi_N^k} \right]^T . \] 
{\em Periodic boundary conditions} are used throughout, while the SLiMM is initialized 
at $\tau=0$ with a single-site excitation of amplitude $A_m$. The excited SQUID is 
either of kind $A$, $B$, or $C$. For isotropic coupling between SQUIDs, i.e., for 
$\lambda_x =\lambda_y$, a single-site excitation of either a $B$ or a $C$ SQUID
provides identical results due to symmetry. 

For the identification of the localized states that may be formed either due to the 
flat band or the nonlinearity, and the quantification of their degree of localization, 
two statistical measures will be used; the {\em energetic participation ratio} $P_e$ 
and the two-dimensional {\em second moment} $M_2$, which are given, respectively, by 
\cite{deMoura2003,Laptyeva2012,Mulansky2012}
\begin{equation}
\label{18}
   P_e =\frac{ 1}{ \sum_{n,m} \epsilon_{n,m}^2 } ,
\end{equation}
and
\begin{equation}
\label{19}
   M_2 =\sum_{n,m} \left\{ ( n -\bar{x} )^2 +m -\bar{y} )^2 \right\} \epsilon_{n,m} ,
\end{equation}
where $\epsilon_{n,m} =H_{n,m}/H$ is the normalized energy density, and $\bar{x}$, 
$\bar{y}$ are the coordinates of the {\em "center of energy"}
\begin{equation}
\label{20}
  \bar{x} =\sum_{n,m}  n \epsilon_{n,m} , \qquad
  \bar{y} =\sum_{n,m}  m \epsilon_{n,m} .
\end{equation}
Note that $P_e$ measures roughly the number of excited cells in the system; its values
range from $P_e =1$ (strong localization, all the energy in a single cell) to $P_e =N$,
with $N=N_x N_y$ (equipartition of the energy over the $N$ SQUIDs). That measure has 
been also used to quantify the degree of diffraction in Kagom{\'e} photonic lattices 
\cite{Vicencio2014}. The second moment $M_2$ quantifies the squared width of the 
state, hence, its spreading.

Eqs. (\ref{6}) with $\gamma=0$ and $\phi_{ext} =0$, implemented with periodic boundary 
conditions are initialized with single-site excitations of the form
\begin{eqnarray}
\label{21}
  \phi_{n,m}^k (\tau=0) =\left\{ \begin{array}{ll}
         A_m,  & \mbox{if $n=n_e$ and $m=m_e$};    \\
         0,    & \mbox{otherwise ,} \end{array} \right. \\ 
\label{22}
 \dot{\phi}_{n, m}^k (\tau=0) =0 , \mbox{for any $n$, $m$} , 
\end{eqnarray}
where $A_m$ is the amplitude of the initial excitation, and $k=A$, $B$ or $C$. The 
excited SQUID belongs to the unit cell with $n=n_e, m=m_e$, with $n_e =N_x/2$ and 
$m_e =N_y/2$. The SLiMM is initialized with $A_m$ spanning several orders of magnitude, 
and for each $A_m$ several quantities such as the energy, the localization measures, 
and the ratio $r =| H (\tau) -H (0) | / H (0)$ are monitored during temporal evolution. 
Typically, a time-step $h=T_{SQ} /1000$, where $T_{SQ} =2\pi /\Omega_{SQ}$, is used 
in the simulations. However, it has been checked that smaller time-steps provide 
practically identical results. It has been also checked that in all runs the ratio 
$r$ remains less than $5 \times 10^{-6}$ for the time step $h$ above. 
\begin{figure}[!t]
\includegraphics[angle=0, width=0.9 \linewidth]{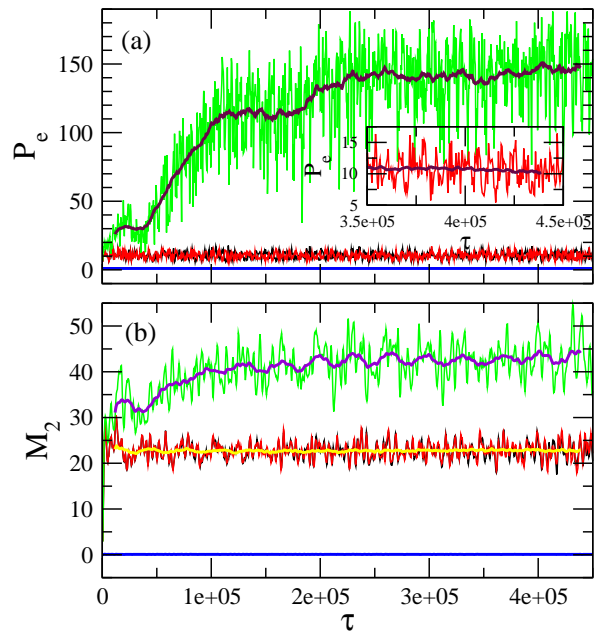} 
\caption{(Color online)
The energetic participation number $P_e$ and the second moment $M_2$ for a SQUID Lieb
lattice with $N_x =N_y =16$, $\lambda_x =\lambda_y =-0.02$, and $\beta_L =0.86$ as a
function of the normalized time $\tau$. The SQUID of kind $C$ (edge) in the 
$(n_e,m_e)-$th cell is initially excited with amplitude $A_m$. 
(a) $P_e$ as a function of $\tau$ for $A_m =10^{-3}$ (black); $10^{-2}$ (red); 
$10^{-1}$ (green); $1$ (blue).
(b) $M_2$ as a function of $\tau$ for $A_m =10^{-3}$ (black); $10^{-2}$ (red); 
$10^{-1}$ (green); $1$ (blue).
Inset: $P_e$ as a function of $\tau$ for $A_m =10^{-2}$ and its running average
over $5000~T_{SQ}$ time units.
\label{fig5}
}
\end{figure}

\section{Flat-Band and Nonlinear Localization}
The typical time-dependence of $P_e$ and $M_2$ when an edge SQUID (i.e., a SQUID $C$) 
is initialy excited with amplitude $A_m$ is shown in Figs. \ref{fig5}a and 
\ref{fig5}b, respectively, for $A_m =0.001$ (black), $0.01$ (red), $0.1$ (green),
and $1$ (blue). Note that the curves for $A_m =0.001$ and $0.01$ almost coincide;
for lower initial amplitudes the results are practically identical to those obtained
for $A_m =0.001$. For such low initial amplitudes the SLiMM remains in the (almost)
linear regime, in which localized flat-band states are expected to be observed.
Indeed, as can be seen in Fig. \ref{fig5}a, as well as in the inset for $A_m =0.01$,
$P_e$ has a running average over $5000~T_{SQ}$ time units which is about eleven 
($P_e \simeq 11$, inset) indicating substantial localization. The existence of a 
localized state is advocated in this case by the corresponding second moment $M_2$, 
which running average over $5000~T_{SQ}$ time units (yellow curve) attains a constant 
value for relatively long integration times ($M_2 \simeq 22$). The constancy of $M_2$ 
is interpreted as the termination of the energy spreading away from the site on which 
it was initially localized.  
\begin{figure}[!t]
\includegraphics[angle=0, width=0.9 \linewidth]{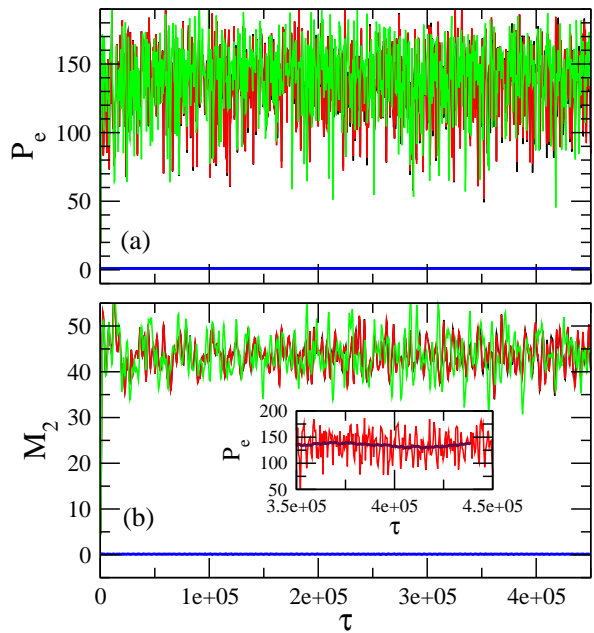} 
\caption{(Color online)
The energetic participation number $P_e$ and the second moment $M_2$ for a SQUID Lieb
lattice with $N_x =N_y =16$, $\lambda_x =\lambda_y =-0.02$, and $\beta_L =0.86$ as a
function of the normalized time $\tau$. The SQUID of kind $A$ (corner) in the 
$(n_e,m_e)-$th cell is initially excited with amplitude $A_m$. 
(a) $P_e$ as a function of $\tau$ for $A_m =10^{-3}$ (black); $10^{-2}$ (red); 
    $10^{-1}$ (green); $1$ (blue).
(b) $M_2$ as a function of $\tau$ for $A_m =10^{-3}$ (black); $10^{-2}$ (red); 
    $10^{-1}$ (green); $1$ (blue).
    Inset: $P_e$ as a function of $\tau$ for $A_m =10^{-2}$ and its running average
    over $5000~T_{SQ}$ time units.
\label{fig6}
}
\end{figure}
\begin{figure*}[!t]
\includegraphics[angle=0, width=0.8 \linewidth]{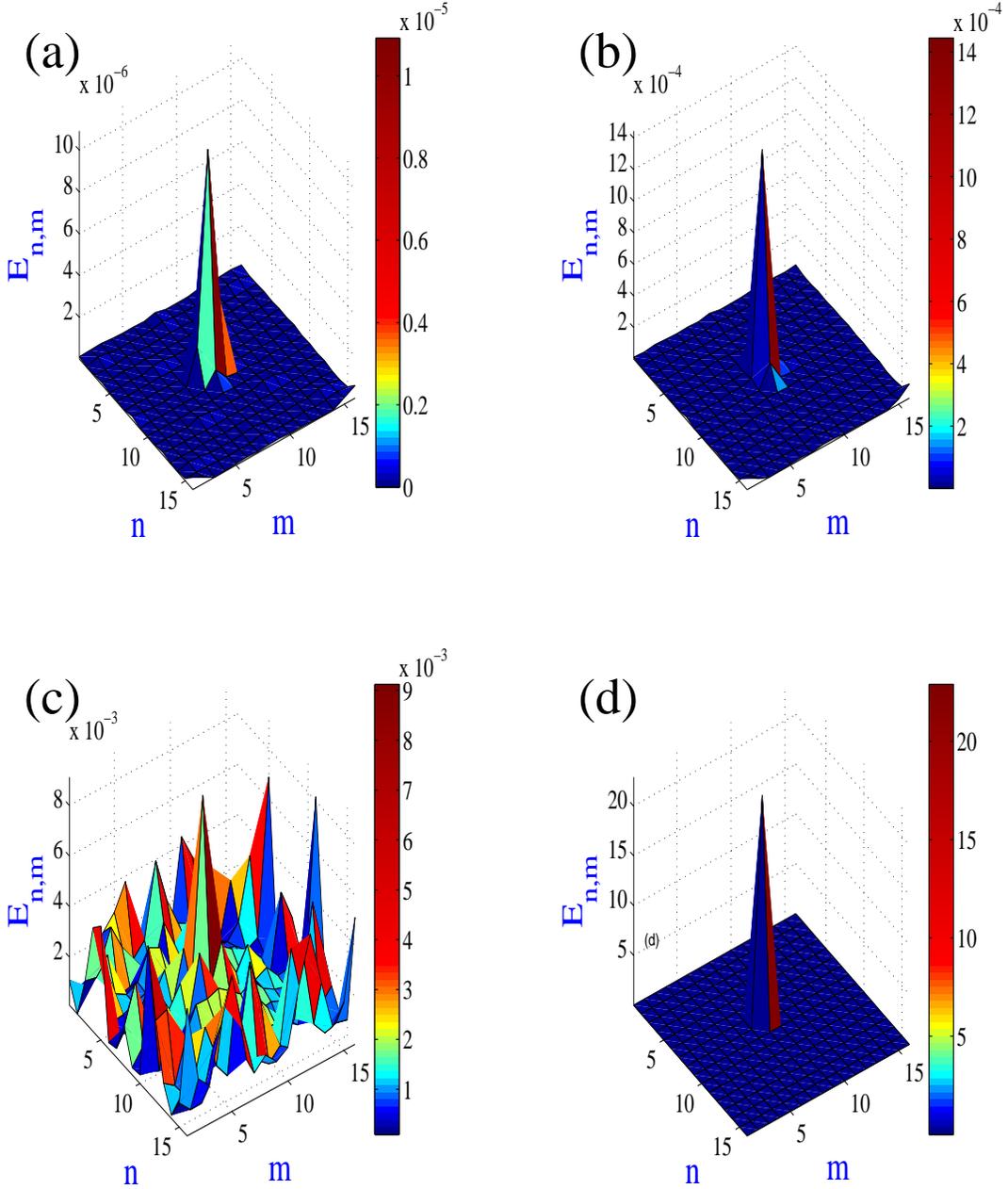} 
\caption{(Color online)
Energy density $E_{n,m} =H_{n,m}$ (energy per unit cell) plotted as a function of $n$
and $m$, after the equations for the SQUID Lieb metamaterial have been integrated 
for $10^5~T_{SQ}$ time units. An edge ($C$) SQUID is initially excited with amplitude
$A_m =0.001$ (a); $0.01$ (b); $0.1$ (c); $1$ (d).
Parameters:  $N_x =N_y =16$, $\lambda_x =\lambda_y =-0.02$, and $\beta_L =0.86$.
\label{fig7}
}
\end{figure*}
For $A_m =0.1$, the inspection of the corresponding (green) curve and its running 
average (maroon) reveals a dramatic change in the behavior of $P_e (\tau)$; the value 
of the latter increases more or less linearly with increasing $\tau$ until it 
saturates at a rather high value around $P_e \sim 140$. Note however the plateaus in 
the running 
average curve which indicate that the SLiMM passes through several metastable states 
until it reaches the steady one. The second moment $M_2$ in this case oscillates around
$43$. Finally, for $A_m =1$ significant nonlinear effects 
come into play that favor strong localization with $P_e \sim 1$; thus, all the energy 
initially provided to the system at a single site, it practically remains there! 
This is actually the reason why the value of $M_2$ remains for all times close to zero 
($M_2 \simeq 0.1$, there is no spreading of energy whatsoever). Clearly, three 
different regimes can be identified; the (almost) linear regime, in which flat-band 
localization is possible, the intermediate regime, in which no localization is observed 
and the initial energy is eventually spread (in time-scales longer than those shown 
here) over the whole lattice, and the nonlinear regime in which localization in the 
form of intrinsically localized modes or discrete breathers, is observed. The size of 
fluctuations, e.g., in the curves for $P_e$, depends on that regime which in turn is 
determined by the initial condition (excitation); thus, fluctuations are weak in the 
linear, strong in the intermediate, and vanishing in the nonlinear regime.      

The corresponding time-dependence of $P_e$ and $M_2$ when a corner SQUID (i.e., a 
SQUID $A$) is initialy excited with amplitude $A_m$ is shown in Figs. \ref{fig6}a 
and \ref{fig6}b, respectively. In this case, there is no localization in the linear 
and the intermediate regimes, i.e., for $A_m =0.001$ (black), $0.01$ (red), and $0.1$ 
(green), as can be inferred by the large values of $P_e$ whose running average over 
$5000 ~T_{SQ}$ time units is about $140$ ($P_e \sim 140$). At the initial stage of 
time integration which is not visible on the scale of the temporal axis of Fig. 
\ref{fig6}, both $P_e$ and $M_2$ have low values; however, within a few thousands time 
units they gradually grow to their high values. Note that the average of the curves 
for $M_2$ ($\sim 43$) is very close to that of the average of the corresponding curve 
for $A_m =0.1$ in Fig. \ref{fig5}b (intermediate regime). For high initial amplitude 
($A_m =1$), however, strong localization due to nonlinearity is again observed. For 
such high values of $A_m$ the localized state which is generated either by initially 
exciting an $A$ or a $C$ SQUID does not reveal any significant difference. When there 
is no localization, the fluctuations of both $P_e$ and $M_2$ are again very strong. 
The results presented in Figs. \ref{fig5} and \ref{fig6} have been obtained for 
$\lambda_x =\lambda_y$, i.e., in the case of an isotropic Lieb lattice in the 
nearest-neighbor approximation. In this case, for single-site initial excitations of 
either a $C$ SQUID or a $B$ SQUID (i.e., of edge SQUIDs), the results are practically
identical.   

\begin{figure}[!h]
\includegraphics[angle=0, width=0.9 \linewidth]{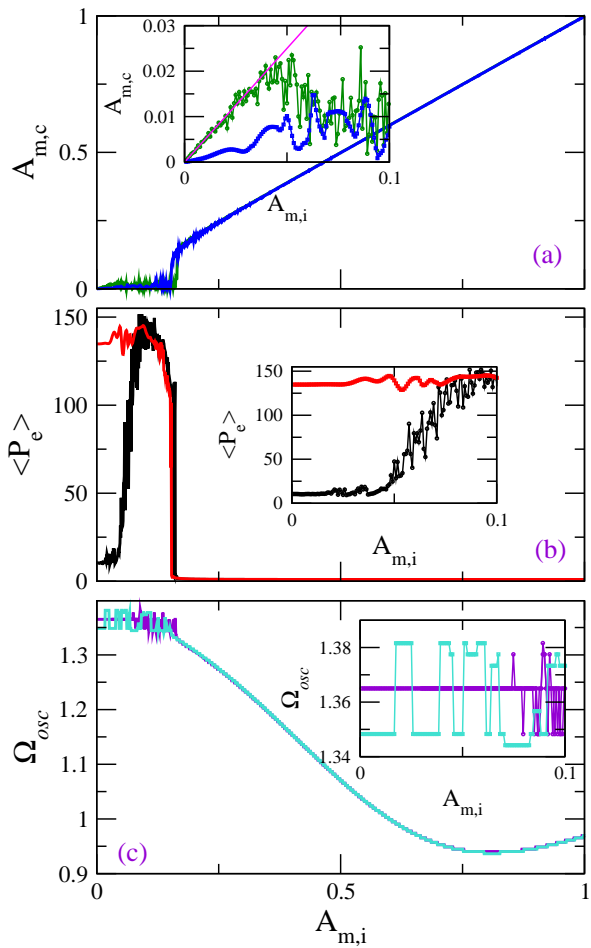} 
\caption{(Color online)
(a) The amplitude $A_{m,c}$ of the flux $\phi_{n_e,m_e}^k$ of the initially excited 
    SQUIDs with $k=A$ (blue curve) and $k=C$ (green curve) calculated at the end of 
    the integration time as a function of the initial excitation amplitude $A_{m,i}$. 
    Inset: Enlargement around low $A_{m,i}$. The line $A_{m,i} /2$ is shown in 
    magenta color. 
(b) The energetic participation ratio averaged over the integration time, $<P_e>$, 
    (transients were discarded) for the SQUID Lieb metamaterial when a corner ($A$) 
    SQUID (red curve) and an edge ($C$) SQUID (black curve) is initially excited, as 
    a function of the initial excitation amplitude $A_{m,i}$. 
(c) The oscillation frequency $\Omega_{osc}$ of the flux $\phi_{n_e,m_e}^k$ of the 
    initially excited $k=A$ (violet curve) and $k=C$ (turquoise curve) SQUID as a 
    function of the initial excitation amplitude $A_{m,i}$.
    Parameters:  $N_x =N_y =16$, $\lambda_x =\lambda_y =-0.02$, and $\beta_L =0.86$.
\label{fig8}
}
\end{figure}

The above scenario is confirmed by inspecting the corresponding energy density plots,
i.e., the plots of the energy density $E_{n,m} =H_{n,m}$ on the $n-m$ plane, which
are shown in Fig. \ref{fig7}. In Fig. \ref{fig7}a, for $A_m =0.001$, the energy density 
$E_{n,m}$ is clearly localized, although not on only one unit cell; the energetic
participation ratio is in this case $P_e \simeq 10.5$. A similar pattern is obtained 
for $A_m =0.01$, as shown in Fig. \ref{fig7}b, in which the maximum of the energy 
density is approximately two orders of magnitude larger than that in Fig. \ref{fig7}a. 
In Fig. \ref{fig7}c, there is clearly no localization, as it can also be inferred by 
the large participation ratio $P_e \simeq 140$. In Fig. \ref{fig7}d, in which the 
localization is due to the nonlinearity, the energy is almost completely localized, 
and $P_e \simeq 1$.    

It should be noted that there are particular types of modes which cannot be 
efficiently excited in the SLiMM with initial single-site excitations used here. 
As an example, consider the application of the constraint $\phi_{n,m}^A =0$ for all 
the SQUIDs of kind $A$. That case has been also considered in a rhombic (quasi-) 
one-dimensional system with three waveguids per unit cell, whose coupling functions 
are the same with those of the equations for the SLiMM \cite{Maimistov2017}. By 
setting $\phi_{n,m}^A =0$ for all $n$ and $m$ and 
$\phi_{n,m}^B =\phi_{n,m}^C =\delta_{n,n_1} \delta_{m,m_1}$,
with $n_1$ and $m_1$ integers, we get from the first of Eqs. (\ref{8}) or the first of
Eqs. (\ref{6}) with $\gamma=0$ and $\phi_{ext} =0$, 
that $\lambda_x \phi_{n_1,m_1}^B =-\lambda_y \phi_{n_1,m_1}^C$ 
($\phi_{n,m}^B =\phi_{n,m}^C =0$ for $n \neq n_1$ and $m \neq m_1$). That particular 
solution for the SLiMM system (either the linearized one or not) certainly cannot be 
obtained using single-site initial excitations.

In order to roughly determine the boundaries between the linear, intermediate, and 
nonlinear regimes, the averages of several quantities over the {\em steady-state 
integration time $\tau_{int}$} were calculated for a wide range of initial excitation 
amplitudes $A_{m}=A_{m,i}$. An edge ($C$) SQUID is initially excited with amplitude 
$A_{m,i}$ and Eqs. (\ref{6}) with $\gamma=0$ and $\phi_{ext} =0$ are integrated in 
time for $\tau_{int} =10^5~T_{SQ}$ time units, to allow for transients to die out 
(the obtained results are discarded) and the steady state to be reached. Then, in the
steady state, the equations are integrated in time for $\tau_{int}$ more time units , 
and the energetic participation ratio averaged over $\tau_{int}$ is calculated. At the 
end of the steady-state integration time, the amplitude of the flux of the excited 
SQUID, $A_{m,c}$, and the oscillation frequency of the flux through the loop of the 
excited SQUID, $\Omega_{osc}$, are also calculated. The same calculations are 
performed for an initially excited corner SQUID $A$, and the results for both cases 
are shown in Fig. \ref{fig8}. In Fig. \ref{fig8}a, the calculated amplitude $A_{m,c}$ 
of the flux $\phi_{n_e,m_e}^k$ through the loop of the SQUID with $k=A$ (blue) and 
$k=C$ (green) is shown along with an enlargement for low $A_{m,i}$ (inset). As it can 
be observed, $A_{m,c}$ attains low values for low initial amplitudes $A_{m,i} < 0.15$, 
while for $A_{m,i} > 0.15$ the calculated amplitude $A_{m,c}$ increases linearly with 
increasing $A_{m,i}$, according to the approximate relation $A_{m,c} \simeq A_{m,i}$. 
The behavior for $A_{m,i} > 0.15$ is a result of the strong 
localization taking place due to nonlinearities and it does not depend on which kind 
of SQUID (edge or corner) is initially excited.
However, a closer look to the two curves for $A_{m,i} < 0.15$, reveals significant 
differences, especially for $A_{m,i} < 0.05$, which can be seen more clearly in the 
inset. In this regime the calculated amplitude $A_{m,c}$ for $k=C$ follows the relation 
$A_{m,c} \simeq A_{m,i}/2$, indicating localization due to the flat band. This 
conclusion is also supported by Figs. \ref{fig8}b and \ref{fig8}c. 
In Fig. \ref{fig8}b, the energetic participation ratio averaged over $\tau_{int}$,
$<P_e>$, for low values of $A_{m,i}$ attains very different values depending on which 
kind of SQUID is initially excited ($A$ or $C$); specifically, while $<P_e> \sim 10.5$ 
for the SLiMM when a $C$ SQUID is initially excited (black), it is $<P_e> \sim 140$ 
when an $A$ SQUID is initially excited (red). That large difference between the values 
of $<P_e>$ is due to flat-band localization in the former case and delocalization in 
the latter case since no flat-band modes are excited. In the inset, it can be observed 
that $<P_e>$ for an initially excited $C$ SQUID starts increasing for $A_{m,i} > 0.05$
indicating gradual degradation of flat-band localization and meets the $<P_e>$ curve
for an initially excited $A$ SQUID at $A_{m,i} \sim 0.1$. In Fig. \ref{fig8}c, 
for $A_{m,i} < 0.15$, the oscillation frequency of the flux through the initially 
excited SQUID $\Omega_{osc}$ (either $A$ or $C$), has a value around that of the 
linear resonance frequency of a single SQUID, $\Omega_{SQ}$ ($\Omega_{SQ} \simeq 1.364$ 
for the parameters of Fig. \ref{fig8}). As it can be seen in the inset, when a $C$ 
SQUID is initially excited (violet), then up to high accuracy 
$\Omega_{osc} =\Omega_{SQ}$ for initial amplitudes up to $A_{m,i} \sim 0.075$. 
However, when an $A$ SQUID 
is initially excited (turquise), the frequency $\Omega_{osc}$ jumps slightly above and 
below $\Omega_{SQ}$ irregularly, but it remains within the bandwidth of the linear
frequency spectrum. For $A_{m,i} > 0.15$, the frequency $\Omega_{osc}$ decreases with 
increasing $A_{m,i}$, although it starts increasing again with increasing $A_{m,i}$ 
at $A_{m,i} \sim 0.8$. In this regime, nonlinear localized modes of the breather 
type are formed, which frequency lies outside the linear frequency spectrum and 
depends on its amplitude, as it should be. From this figure it can thus be inferred 
that flat-band localization occurs for initial amplitudes up to $A_{m,i} \simeq 0.05$ 
(linear regime), while delocalization occurs in the interval $0.05 < A_{m,i} < 0.15$ 
(intermediate regime). For larger $A_{m,i}$, strong nonlinear localization occurs 
(nonlinear regime). This rough estimation for the boundaries between the three regimes 
is of course parameter dependent. Remarkably, flat-band localization occurs only when 
an edge SQUID ($B$ or $C$) is initially excited. The excitation of a corner ($A$) 
SQUID does not lead to excitation of flat-band modes and thus such a localized initial 
state rapidly delocalizes. On the other hand, the observed flat-band localization is 
not very strong as compared to the nonlinear localization. This is probably due to 
the fact that a single-site excitation of a $B$ or $C$ SQUID does not correspond to 
an exact localized flat-band eigenmode.   

\begin{figure}[!h]
\includegraphics[angle=0, width=0.9 \linewidth]{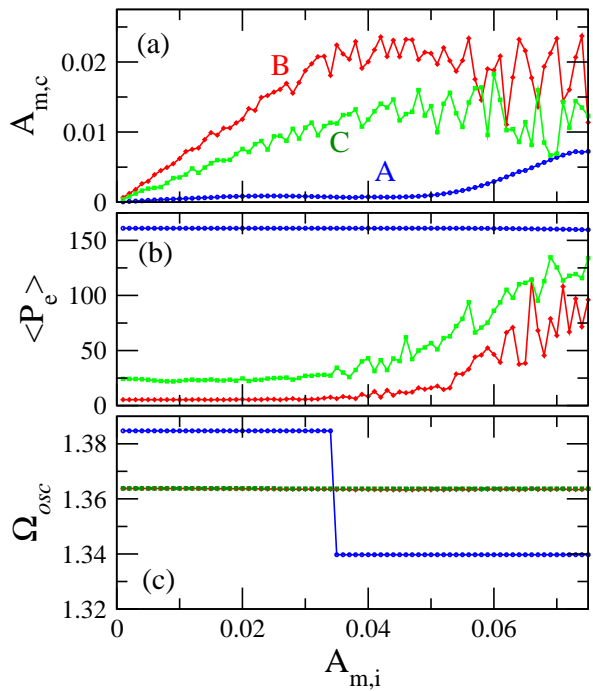} 
\caption{(Color online)
(a) The amplitude $A_{m,c}$ of the fluxes $\phi_{n_e,m_e}^k$ of the excited SQUIDs 
    with $k=A$ (black), $B$ (red), and $C$ (green), in the $(n_e,m_e)$th unit cell, 
    calculated at the end of the integration time as a function of relatively low 
    initial excitation amplitudes $A_{m,i}$;
(b) the corresponding energetic participation ratio $i<P_e>$ averaged over the 
    steady-state integration time $\tau_{int}$ (transients were discarded), and  
(c) the corresponding oscillation frequency $\Omega_{osc}$ of the fluxes 
    $\phi_{n_e,m_e}^k$, for $N_x =N_y =16$, $\beta_L =0.86$, and anisotropic coupling 
    $\lambda_x =-0.02$, $\lambda_y =-0.03$.
\label{fig9}
}
\end{figure}

In the case of anisotropic coupling, i.e., for $\lambda_x \neq \lambda_y$, single-site 
excitations of $B$ and $C$ SQUIDs give different results as expected due to the 
lowering of symmetry \cite{Guzman-Silva2014}. Typical curves for the amplitude of the 
flux $\phi_{n_e,m_e}^k$ of the excited SQUID $A_{m,c}$ ($k=A$, $B$, and $C$), the 
energetic participation ratio averaged over the steady-state integration time $<P_e>$, 
and the oscillation frequency $\Omega_{osc}$ of the flux through the excited SQUID, 
are shown in Fig. \ref{fig9} for anisotropic nearest-neighbor coupling, 
$\lambda_y =1.5\, \lambda_x =-0.03$, as a function of relatively low initial 
excitation amplitudes $A_{m,i}$ for which flat-band localization is expected. As can 
be observed in Fig. \ref{fig9}a, flat-band localization occurs when either of the edge 
SQUIDs are excited with $A_{m,i} < 0.04$. For larger values of $A_{m,i}$, localization 
starts degrading as it can be confirmed from the corresponding curves of $<P_e>$ in 
Fig. \ref{fig9}b. Here, it is also apparent that initial excitations of $B$ and $C$ 
SQUIDs do not lead to a state with the same degree of localization; indeed, $<P_e>$ is 
respectively $\sim 5$ and $\sim 25$ (while the corresponding $<P_e>$ for initial 
excitations of an $A$ SQUID is about $160)$. The corresponding oscillation frequencies 
of the fluxes in the case of $B$ or $C$ SQUID initial excitations are practically 
equal to that of the single SQUID resonance $\Omega_{SQ} \simeq 1.364$. For initial 
excitations of an $A$ SQUID, the oscillation frequency is very close to that of either 
the upper or the lower boundary of the linear frequency spectrum.  

\section{Conclusions}
The dynamic equations for the fluxes threading the SQUID loops of a driven-dissipative
SLiMM have been derived, along with the corresponding linear frequency spectrum. The 
Lieb lattice geometry results in a spectrum with two dispersive bands, which form a 
Dirac cone at the corners of the first Brillouin zone, and a flat band crossing those 
Dirac points. The localization properties of Hamiltonian SLiMMs, i.e., those without 
dissipation and driving terms, have been determined through numerical simulations for
single-site initial excitations of varying amplitude. Flat-band localization, i.e., 
the emergence of localized flat-band states, is observed when an edge ($B$ or $C$) 
SQUID of the unit cell of the SLiMM is initially excited with low amplitude. To the 
contrary, no such states are generated when a corner ($A$) SQUID of the unit cell of 
the SLiMM is initially excited with low amplitude. These results are compatible with 
the experiments on photonic Lieb lattices \cite{Mukherjee2015a}. For sufficiently 
high amplitude of the initial excitation of either a corner or an edge SQUID, 
localization due to nonlinearities in the form of discrete breathers is observed. The 
linear (low amplitude initial excitations) and the nonlinear regimes (high amplitude 
initial excitations), in which flat-band localized states and discrete breathers, 
respectively, can be generated, are separated by an intermediate regime in which 
neither type of localization is observed. This dynamic behavior is quite different 
from that observed in, e.g., two-dimensional Kagom{\'e} lattices in which families of 
nonlinear localized modes in the form of discrete solitons or discrete breathers may 
bifurcate from localized linear modes of the flat band 
\cite{Vicencio2013,Johansson2015}. 
Here, relatively high-amplitude initial excitations ($A_{m,i} > 0.05$) excite nonlinear 
effects in the SQUIDs which destroy the flatness of the flat-band which has been 
obtained in the linear limit. At the same time, however, these nonlinear effects are 
not strong enough to help the initial excitation to remain localized (self-trapped); 
that occurs only when the amplitude of the initial excitation exceeds a particular, 
parameter-dependent threshold ($A_{m,i} \simeq 0.15$ for the parameters of Fig. 
\ref{fig8}).   

\section*{ACKNOWLEDGMENT}
This work is partially supported by the Ministry of Education and Science of the 
Russian Federation in the framework of the Increase Competitiveness Program of NUST 
"MISiS" (No. K2-2015-007) and by the Ministry of Education and Science of the Republic 
of Kazakhstan (Contract $\#$ 339/76 --2015). 
NL gratefully acknowledges the Laboratory for Superconducting Metamaterials, NUST 
"MISiS" for its warm hospitality during visits.




\end{document}